# Polarized and bright telecom C-band single-photon source from InP-based quantum dots coupled to elliptical Bragg gratings

Zhenxuan Ge,[1,2] Tunghsun Chung,[2,3] Yu-Ming He,[1,2,3,*] Mohamed Benyoucef,[4,*] and Yongheng Huo[1,2,3,*]

[1]Hefei National Research Center for Physical Sciences at the Microscale and School of Physical Sciences, University of Science and Technology of China, Hefei 230026, China

[2]Shanghai Research Center for Quantum Science and CAS Center for Excellence in Quantum Information and Quantum Physics, University of Science and Technology of China, Shanghai 201315, China

[3]Hefei National Laboratory, University of Science and Technology of China, Hefei 230088, China

[4]Institute of Nanostructure Technologies and Analytics (INA), Center for Interdisciplinary Nanostructure Science and Technology (CINSaT), University of Kassel, Heinrich-Plett-Str. 40, 34132 Kassel, Germany

**ABSTRACT:** Bright, polarized, and high-purity single-photon sources in telecom wavelengths are crucial components in long-distance quantum communication, optical quantum computation and quantum networks. Semiconductor InAs/InP quantum dots (QDs) combined with photonic cavities provide a competitive path leading to optimal single-photon sources in this range. Here, we demonstrate a bright and polarized single-photon source operating in the telecom C-band based on an elliptical Bragg grating (EBG) cavity. With a significant Purcell enhancement of $5.25 \pm 0.05$, the device achieves a polarization ratio of 0.986, single-photon purity of $g^2(0) = 0.078 \pm 0.016$ and single-polarized photon collection efficiency of ~24% at the first lens (NA=0.65) without blinking. These findings suggest that C-band QD-based single-photon sources are potential candidates for advancing quantum communication.

Bright, polarized, and high-purity single-photon sources are essential elements in long-distance quantum communication[1-4] and optical quantum computation.[5-7] Semiconductor quantum dots (QDs), due to their compatibility with existing semiconductor processes, have received long-term attention in realizing solid-state optimal single-photon sources. In recent years, QDs combined with state-of-the-art photonic cavities have achieved remarkable progress in brightness,[8,9] single-photon purity,[10,11] indistinguishability,[9,12,13] and entanglement fidelity.[14–16] However, the spectral range (780 and 900 nm) of these achievements falls outside the telecom C-band, resulting in high photon loss and dispersion over long-distance fiber transmission. Using quantum frequency conversion is one practical option to convert the non-C-band photons into C-band, but it will bring an additional 50% photon efficiency loss.[17]

To overcome this challenge, researchers have employed two types of QDs operating at telecom wavelengths. The first is InAs/InP QDs,[18-23] which can emit telecom-wavelength photons naturally due to a smaller lattice mismatch between In(Ga)As and InP than InAs/GaAs. The second is InAs/InGaAs/GaAs QDs,[24,25] where a metamorphic buffer layer (MMB) is deployed to reduce the lattice mismatch and push the QD emission toward the telecom wavelength. Various impressive devices based on these two systems have been reported in telecom wavelengths, such as the horn structure,[26] photonic crystals,[27–29] nanobeams,[30] and circular Bragg gratings (CBGs).[31–33] Among them, the horn structure source, with a photon collection efficiency of 11%, successfully achieved quantum key distribution (QKD) in 120 km,[34] verifying the application potential of C-band QDs in quantum communications. Recently, the reported state-of-the-art CBG cavity reaches a collection efficiency of 17.4% in the telecom C-band, with a low $g^2(0) = 0.0052$,[35] making a step toward realizing brighter and purer C-band single photon sources. In addition to the brightness and purity, a well-defined polarization state of the emitted photons is also essential for many quantum information applications, such as the polarization-based QKD[36] and the implementation of an optical controlled-NOT gate.[37,38] Unfortunately, conventional CBG has no photon polarization selectivity. For nonpolarized single-photon emitters, postselecting one linear polarization state would cause an additional 50% photon loss.

To solve this problem and simultaneously explore the realization of a brighter single-photon source, we demonstrate EBGs with InAs/InP QDs. The elliptical effect has been proven in splitting the cavity mode and increasing the linearly polarized photon collection efficiency in 900 nm QD devices.[39] In the telecom C-band, our EBG shows a 13.23 nm mode splitting, and two individual modes can be clearly separated. The emission of the QD coupled to one cavity mode exhibits high linear polarization with a polarization ratio of 0.986 and the single photon nature with a $g^2(0) =$ $0.078 \pm 0.016$. Meanwhile, the highest single-photon collection efficiency of ∼24% is achieved at saturation excitation with a $g^2(0) = 0.114 \pm 0.022$ without blinking. These results demonstrate the potential of the elliptical Bragg grating in achieving bright, polarized, and high-purity telecom-wavelength single-photon sources.

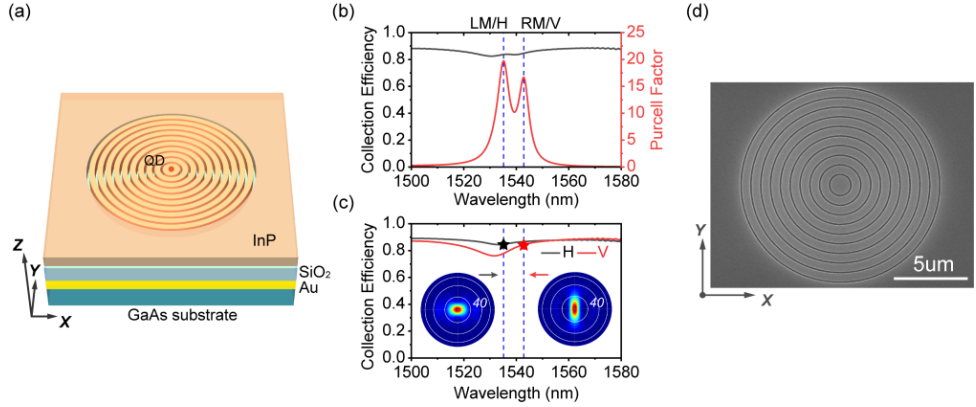

**Figure 1.** (a) Schematic diagram of the EBG. (b) Calculated photon collection efficiency and Purcell factor of the device when the dipole orientation is set at 45 degrees in-plane with respect to the X-axis. The blue dashed lines mark the positions of the left mode (LM/H) and right mode (RM/V). (c) Calculated photon collection efficiency when the dipole orientation is set at 0 degree (H) and 90 degrees (V), respectively. The inset shows the far-field patterns of two orientations at corresponding center wavelengths (blue dashed lines) with NA=0.65 marked. (d) Top view of a completed EBG.

To design the EBG in the telecom C-band, we perform finite-difference time-domain (FDTD) simulations. The structure design is shown in Figure 1a. From top to bottom, the first layer is a 310 nm-thick InP epilayer with InAs QDs embedded in the center, the second layer is a 300 nm-thick $SiO_2$ dielectric layer, the third layer is a 200 nm-thick Au reflector and the last layer is a GaAs substrate. The cavity is defined at the InP layer with the radius of the central disk set at 660 nm, the grating period set at 660 nm, the trench set at 120 nm, and the aspect ratio (Y: X) fixed at 0.99.

First, we set the dipole orientation at 45 degrees with respect to the X-axis. Figure 1b shows a clear spectrum splitting for 7.7 nm, indicating a successful spectral separation of modes LM/H (left mode, horizontal polarization) and RM/V (right mode, vertical polarization) with their corresponding center wavelengths 1535.15 nm and 1542.85 nm, respectively. Photon collection efficiency (>80%) reaches a plateau in the range 1500-1580 nm, enough to cover the full telecom C-band.

Then, we set the dipole orientation at 0 and 90 degrees, aiming to fully excite one of the two modes, respectively. The calculated photon collection efficiency is shown in Figure 1c. At the spectral positions of LM/H and RM/V, efficiencies of 84.7% and 84.5% are achieved (marked by the black star and red star), respectively. The far-field patterns at these two wavelengths, oriented along the X and Y axes, both exhibit a clear Gaussian-like energy concentration within NA=0.65, which is beneficial for objective collection and fiber coupling (the inset in Figure 1c).

The sample is grown by a molecular beam epitaxy (MBE) system on an InP epi-ready substrate.[22,28] A 100 nm InP buffer layer is first grown on the substrate, followed by a 1400 nm InGaAs sacrificial layer. A 155 nm-thick InP layer is then grown, followed by InAs QDs, which are grown by using a ripening technique,[20,21] allowing the control of density and emission wavelength. To cover the QDs, a 155 nm-thick InP capping layer is grown. And the device is fabricated through a flip-chip Au-Au bonding process,[40] E-beam lithography, and two steps of dry etching (more details are provided in the Supporting Information). Figure 1d shows a scanning electron microscopy (SEM) image of a typical EBG on the device.

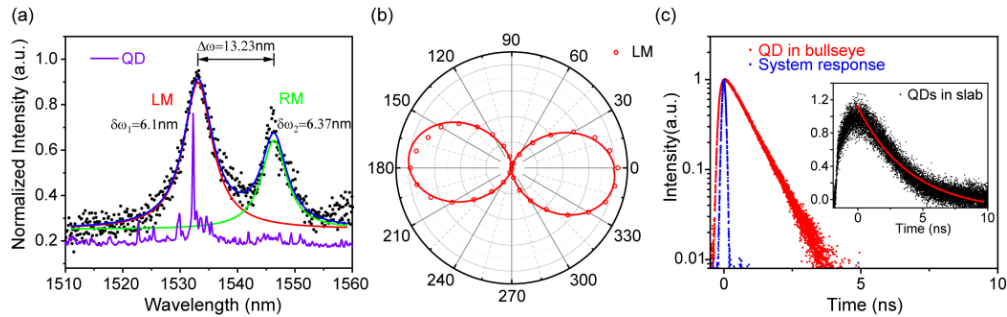

**Figure 2.** (a) Normalized spectrum of the investigated QD; the cavity mode splits into LM and RM. (b) The polarization of the QD emission in the mode LM. (c) The lifetime of the investigated QD inside the cavity. A comparison decay time of the QD in the slab is shown in the inset.

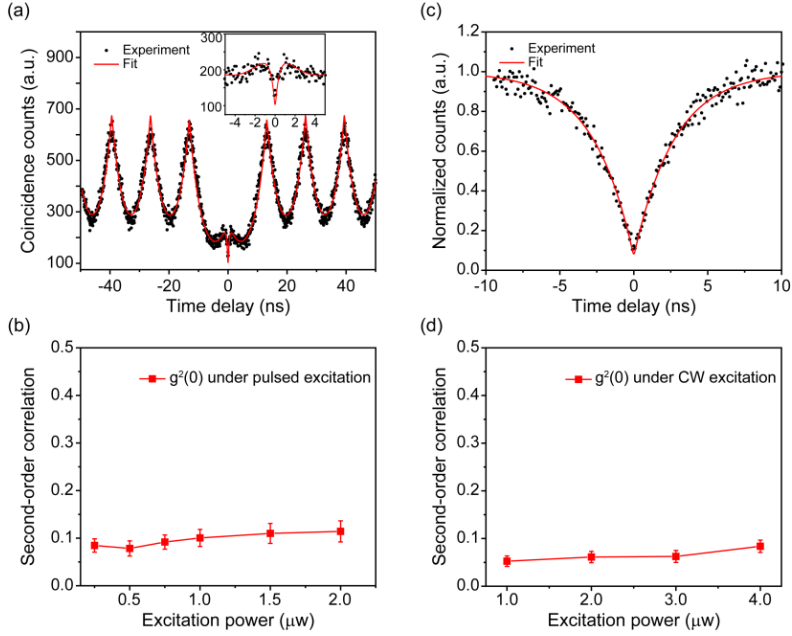

**Figure 3.** (a) The second-order correlation of the investigated QD under pulsed excitation at saturation power. The inset shows the antibunching dip near zero-time delay. (b) The $g^2(0)$ values at different excitation powers under pulsed excitation. (c) $g^2(\tau)$ under continuous wave (CW) excitation. (d) $g^2(0)$ values at different CW excitation powers.

With a confocal microscope setup, optical characterization is carried out when the sample is cooled to 4 K in a cryostat. An 890 nm pulsed laser (excitation repetition rate 76 MHz) and a white light source are used to excite QDs and the cavity modes, respectively. The signal photons are collected by a built-in objective lens with NA=0.65 and filtered out by a monochromator.

The spectrum of the cavity is shown in Figure 2a. Due to the elliptical design, the cavity mode splits into two discrete modes, LM and RM, as predicted. Both cavity modes have a broad bandwidth greater than 6 nm, and the 13.23 nm spectral spacing between the two modes is larger than the simulation results, indicating a smaller aspect ratio for the fabricated cavity. A clear QD signal at 1532 nm can be recognized in Figure 2a matching the LM mode, indicating effective QD-cavity coupling. We perform a polarization measurement to characterize the photons emitted by the QD coupled to LM, as shown in Figure 2b. By defining a polarization ratio as $(I_{max} - I_{min})/(I_{max} + I_{min})$, where $I_{max}$ and $I_{min}$ are the maximum and minimum photon emission intensities, respectively, a polarization ratio of 0.986 is obtained. This result confirms that the QD emission has been successfully induced and filtered to a linear polarization by the cavity mode LM.

To investigate the Purcell enhancement of the cavity, we conduct time-resolved photoluminescence measurements to determine the QD lifetime, as shown in Figure 2c. To provide a high temporal resolution, superconducting nanowire single photon detectors (SNSPDs) are used. We use an exponential function convolved with the system response to fit the decay in Figure 2c, and finally extract a lifetime of $0.775 \pm 0.001$ ns for the QD coupled into the cavity. We need to point out that the actual lifetime of the investigated QD may be shorter due to the additional relaxation process caused by the used above-band excitation. As a comparison to the lifetime in the cavity, the lifetimes of several QDs in the slab without an EBG cavity are also measured with the same setup. The inset of Figure 2c shows a typical decay of the QD emission in the slab area, with a time constant of $4.065 \pm 0.042$ ns. This relatively long lifetime could be due to the weaker electron confinement in the InP matrix[41,42] and the reduction of the local density of states of the light field.[43] Compared to the QD in the EBG, we can obtain a Purcell factor of $5.25 \pm 0.05$, clearly revealing effective QD-cavity coupling. Due to the additional relaxation process caused by above-band excitation, we expect the measured lifetimes of the investigated QDs be shorter, especially for the QD in the cavity, and the Purcell factor can be higher after using more precise excitation methods, such as the p-shell or s-shell scheme.

In the evaluation of single-photon sources, single-photon purity is a key parameter. To measure the single-photon purity, we use a Hanbury-Brown and Twiss (HBT) measurement setup. The investigated QD is excited by an 890 nm pulsed laser at saturation power. The second-order autocorrelation measurement result is shown in Figure 3a. Each peak is fitted by a two-sided exponential function and the central dip (inset of Figure 3a) is considered as well.[44] Near zero-time delay, the coincidence counts hit the minimum, showing apparent antibunching but with two obvious shoulders in the vicinity. It indicates the presence of refilling,[31] which means that after the exciton recombines and emits a photon, the carriers in the wetting layer or in other charge trap states nearby will be recaptured by the QD resulting in secondary photon emission, thus increasing the coincidence counts near zero-time delay.[44] After dividing the area around zero-time delay by the average area of nearby peaks, we obtain a raw $g^2(0)$ of 0.152. By extracting the dark counts, we finally obtain a $g^2(0) = 0.114 \pm 0.022$. The remaining multiphoton events can mainly be attributed to the refilling of the QD. To further explore the power dependence of the refilling process, the laser power is swept from a few hundred nW to saturation. The obtained $g^2(0)$ values are displayed in Figure 3b. By gradually decreasing laser power, the $g^2(0)$ value decreases to reach the lowest of $0.078 \pm 0.016$ at a laser power of 0.5 μW. As the excitation power is reduced further, a significant decrease in photon counts occurs, resulting in an unavoidable decline in the signal-to-noise ratio, which causes an increase in $g^2(0)$ values.

To further verify the single-photon purity of this QD, a correlation measurement with a 780 nm CW laser excitation is performed. The excitation power is also set at saturation first, and the result is shown in Figure 3c, where we can obtain a $g^2(0) = 0.084 \pm 0.013$. Figure 3d shows the $g^2(0)$ values with varying the laser power. The best single-photon purity ($g^2(0) = 0.053 \pm 0.011$) is achieved at the lowest power, suggesting that due to our lack of suitable C-band lasers, the non-resonant excitation scheme we now have to use limits the further improvement of the single-photon purity of the investigated QD.

The photon collection efficiency is another key element, which characterizes the brightness of a single-photon source. To evaluate the improvement in photon collection efficiency provided by our EBG, we measure the blinking of the investigated QD through a long-time scale HBT measurement under 76 MHz pulsed excitation under a saturation power. To speed up the data collection, we open the slit of the monochromator wider to allow more signals to pass through. Although this will introduce some noise, it will not affect the measurement result of blinking. As depicted in Figure 4a, it is evident that the coincidence counts remain constant over time, revealing a blinking-free condition for the investigated QD. Therefore, the efficiency loss in blinking can be ignored. By gradually increasing the laser power, the fiber-coupled photon count rate reaches its maximum value of 2.427 M/s at an excitation power of 2.0 μW, as shown by the blue line in Figure 4b, where the photon counts have been corrected by detection efficiency (∼85%). By counting the optical setup efficiency (∼13%), single-photon purities and photon count rates at each power point, we can extract the first lens single-photon collection efficiency at each power point, as shown by the red line in Figure 4b (more details in Supporting Information). The maximum collection efficiency reaches ∼24%. In addition, considering the difference between the fabricated structure and the designed one, the photon collection efficiency of this device could be further improved through fabrication optimizations (see Supporting Information).

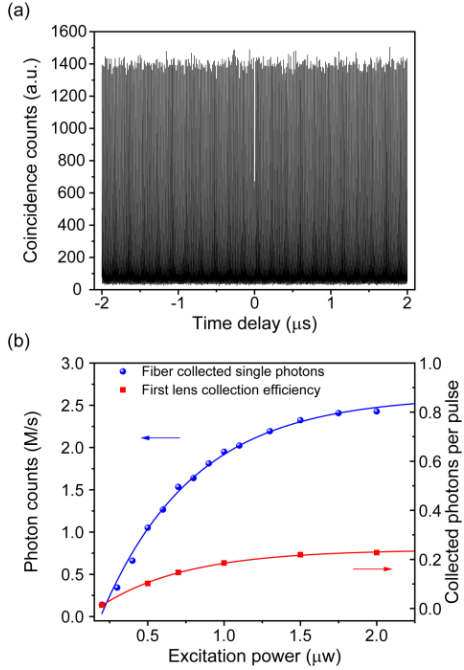

**Figure 4.** (a) Long-time scale second-order autocorrelation measurement shows that coincidence counts do not decrease with time, revealing a blinking-free condition. (b) The fiber collected photon counts and first lens collection efficiency as a function of excitation power under above-band excitation. The photon counts have been corrected by detection efficiency.

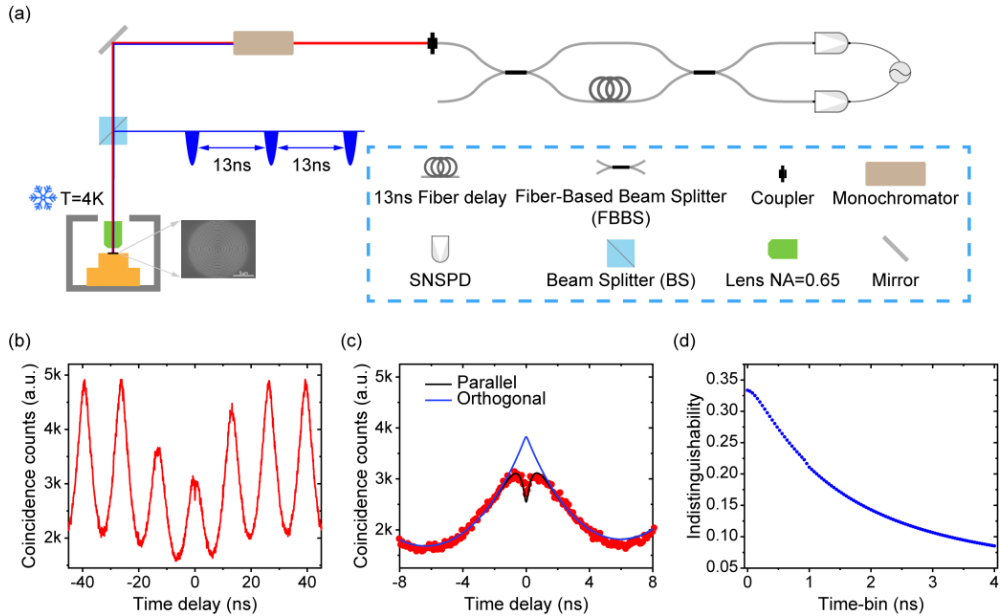

**Figure 5.** (a) Schematic of the indistinguishability measurement setup: an unbalanced fiber-based Mach–Zehnder interferometer with a 13 ns path difference. (b) The measurement result of the parallel setup; a clear dip can be observed at zero-time delay. (c) Comparison between the parallel case and orthogonal case. (d) The indistinguishability as a function of the time-bin shows a post-selected maximum two-photon interference (TPE) visibility, $V_{TPI} = 33.36\% \pm 0.01\%$.

In addition to brightness and single-photon purity, the photon indistinguishability is also required in many quantum information applications, such as the optical controlled-NOT gate,[37] photonic cluster state[45] and entanglement swapping.[46] Here, we use an unbalanced fiber-based Mach–Zehnder interferometer[47] (shown in Figure 5a) to measure the indistinguishability of the emitted photons. The QD is excited by a 76 MHz pulsed laser at saturation power to obtain a maximum photon count rate. The fluorescence photons are filtered and coupled into the interferometer, where one arm of the interferometer is precisely tuned to compensate for the 13 ns delay introduced by the pulse repetition rate. Figure 5b displays the correlation histogram with a parallel polarized configuration. At zero-time delay, a clear dip can be observed, revealing the quantum interference of the emitted photons. For a more obvious and precise comparison, we used a two-sided exponential function convolved with a Gaussian distribution to fit the data in the parallel condition near zero-time delay[27] in Figure 5c (details in Supporting Information). The orthogonal curve is derived from the data fitting since our detectors have a polarization response.[47] By summing the coincidence counts of parallel and orthogonal cases ($S_\parallel$ and $S_\perp$) within different time-bins near zero delay, the visibility of two-photon interference (TPI) with different time-bins can be calculated through $V_{TPI} = 1 - S_\parallel/S_\perp$, as shown in Figure 5d. With a reduction in the time-bin, we finally obtained a maximum indistinguishability value of $V_{TPI} = 33.36 \pm 0.01\%$. By considering the single-photon purity and splitting ratio of our BS (Reflection : Transmission = 54 : 46), the corrected visibility[48] is $V_{corrected} = 43.2 \pm 1.5\%$, which is consistent with the results reported for telecom QDs.[27,35] We attribute this imperfect indistinguishability to the refilling process,[31] induced by charge defects, and the timing jitter[49] caused by our off-resonant excitation. Some recent works have reported several methods that can be used to further improve the performance, such as applying an external electrical field[50,51] to stabilize the charge environment and using p-shell excitation[30,31,52] to excite the QD energy level more precisely.

We have demonstrated and characterized a polarized and bright telecom C-band single-photon source from InP-based QDs coupled to EBG cavities. High linearly polarized photon emission with a polarization ratio of 0.986 was achieved. A significant cavity Purcell enhancement of $5.25 \pm 0.05$ and blinking-free condition were realized. We finally obtained a ∼24% single-photon collection efficiency. Our improved collection efficiency and polarization purity can enhance quantum communications, such as increasing QKD secure key rate to more than twice the previously reported result.[34] Moreover, the photon collection efficiency can be further improved through fabrication optimizations, and a higher excitation repetition rate could be applied to obtain a higher photon count rate. Furthermore, we could control the transition more precisely by employing p-shell[30,31,52] or resonant[53] excitation schemes to suppress the influence of refilling, and the timing jitter caused by the relaxation process. As a result, we can achieve a higher Purcell factor and improve the single-photon purity as well as indistinguishability. Using an external electrical field,[32,50,51] such as a p-i-n diode, is another approach to stabilize the charge environment and control the exciton state, which can further improve the performance of photon sources. Recently, the other EBG work has been reported in telecom O-band,[54] indicating that the polarized telecom-wavelength single-photon source has attracted more attention than before. We believe that our results demonstrate the potential of the InAs/InP QD-based device for achieving an optimal telecom C-band single-photon source, which is crucial for applications in large-scale quantum communication.

## ASSOCIATED CONTENT

**Supporting Information**.
Sample fabrication process, details in collection efficiency analysis, additional simulation results for photon collection efficiency analysis, fitting and correction for indistinguishability. This material is available free of charge via the Internet at http://pubs.acs.org.

## AUTHOR INFORMATION

### Corresponding Author

Yongheng Huo - Hefei National Research Center for Physical Sciences at the Microscale and School of Physical Sciences, University of Science and Technology of China, Hefei 230026, China; Shanghai Research Center for Quantum Science and CAS Center for Excellence in Quantum Information and Quantum Physics, University of Science and Technology of China, Shanghai 201315, China; Hefei National Laboratory, University of Science and Technology of China, Hefei 230088, China. Email: yongheng@ustc.edu.cn

Mohamed Benyoucef - Institute of Nanostructure Technologies and Analytics (INA), Center for Interdisciplinary Nanostructure Science and Technology (CINSaT), University of Kassel, Heinrich-Plett-Str. 40, 34132 Kassel, Germany. Email: m.benyoucef@physik.uni-kassel.de

Yu-Ming He - Hefei National Research Center for Physical Sciences at the Microscale and School of Physical Sciences, University of Science and Technology of China, Hefei 230026, China; Shanghai Research Center for Quantum Science and CAS Center for Excellence in Quantum Information and Quantum Physics, University of Science and Technology of China, Shanghai 201315, China; Hefei National Laboratory, University of Science and Technology of China, Hefei 230088, China. Email: yuminghe@mail.ustc.edu.cn

Authors
Zhenxuan Ge - Hefei National Research Center for Physical Sciences at the Microscale and School of Physical Sciences, University of Science and Technology of China, Hefei 230026, China; Shanghai Research Center for Quantum Science and CAS Center for Excellence in Quantum Information and Quantum Physics, University of Science and Technology of China, Shanghai 201315, China

Tunghsun Chung - Shanghai Research Center for Quantum Science and CAS Center for Excellence in Quantum Information and Quantum Physics, University of Science and Technology of China, Shanghai 201315, China; Hefei National Laboratory, University of Science and Technology of China, Hefei 230088, China.

Author Contributions
Y.H.H, M.B and Y.M.H supervised this project. M.B provided MBE samples. Z.X.G and T.H.C fabricated the devices. Y.M.H and Z.X.G performed the optical characterizations and analyzed the data. Z.X.G completed the manuscript.

Funding Sources
National Natural Science Foundation of China (12012422); The National Key R&D Program of China (2019YFA0308700); The Chinese Academy of Sciences, the Anhui Initiative in Quantum Information Technologies, the Shanghai Municipal Science and Technology Major Project (2019SHZDZX01); The Innovation Program for Quantum Science and Technology (2021ZD0300204, 2021ZD0301400); The Youth Innovation Promotion Association of CAS and DFG (Heisenberg grant-BE 5778/4-1).

Notes
The authors declare no conflicts of interest.

## ACKNOWLEDGMENT

We gratefully acknowledge the financial support from the Natural Science Foundation of Shanghai and the support from the USTC Center for Micro- and Nanoscale Research and Fabrication. We acknowledge Prof. Chaoyang Lu and Prof. Jianwei Pan for help in the project initialization. We acknowledge Runze Liu and Yukun Qiao for technical discussions. Similarly, we acknowledge Andrei Kors for his assistance in the MBE growth process, Johann Peter Reithmaier for his discussion, and Dirk Albert for his technical assistance.

TOC Graphic

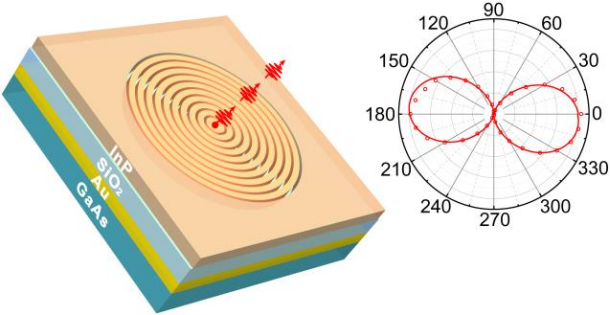